# RECENT DEVELOPMENTS IN COMPUTER MODELING OF AMORPHOUS MATERIALS

**D. A. Drabold, P. Biswas, D. Tafen and R. Atta-Fynn**

*Department of Physics and Astronomy*

*Ohio University*

*Athens, Ohio 45701 USA*

## Introduction

In this paper, we review some recent work on amorphous materials using current "first principles" electronic structure/molecular dynamics techniques. The main theme of the paper is to emphasize new directions in the use of such *ab initio* methods. Some of these, being quite new, need development, but we believe have promise for solving new and important kinds of problems in the physics of glassy and amorphous materials.

Initially, we discuss first principles calculations in broad outline and comment on the various approximations in common use. Then, we describe methods for forming a computer model of amorphous materials. This is an area of intense activity and methods beyond the obvious "quench from the melt" method are showing promise and utility. In this paper, we discuss a new method: "Decorate and Relax", and a new implementation of the Reverse Monte Carlo method. Finally, we discuss the computation of electronic properties, especially carrier transport and time evolution of electron states.

## Hamiltonians, density functionals, and all that

Up until the eighties, dynamic simulation of materials "molecular dynamics" was always based upon an empirical interatomic potential. The idea was to concoct an energy functional that took as input the positions of atoms and as output gave the total energy of the conformation and its gradients (the interatomic forces)[1]. Despite heroic efforts, these potentials always seem to have limited validity -- as they are typically obtained from fitting to a set of experiments, their reliability is better when they are used for structures close to conformations sampled in the fitting process[2]. Since the very structure of amorphous materials is itself a difficult question, such potentials are typically not *designed* using amorphous structures for fitting, so that the utility of empirical potentials is always questioned for amorphous materials. Still, there are important questions in the theory of amorphous materials (which are inaccessible to other more accurate but more expensive methods) in which the use of empirical potentials is valuable. To get beyond empirical potentials, one has to formulate a more fundamental approach to the interatomic forces – which ultimately must involve accurate and parameter-free calculations of the electronic structure of the material. This is usually called "first principles" or *ab initio* modeling. Current *ab initio* modeling of materials is a development arising from some key ideas:

## Density functional theory.

In a celebrated pair of papers, Kohn, Hohenberg and Sham[3] showed that it was possible in principle to exactly compute the ground state properties of an interacting inhomogeneous electron gas, as encountered in molecules or solids. In the first paper, two disarmingly simple proofs were given that 1) the energy of the electronic ground state is a functional of the electronic charge and 2) when this functional is minimized, one has the physical charge density associated with the ground state and the energy is the exact energy. This result is important, since it provides license to link ideas of electronic energy and charge (rather than bewildering many-body wavefunction which is exceedingly difficult to compute, for a complicated system even for the ground state). However, the results would have primarily graced the pages of textbook were it not for the second paper, which showed how to estimate the quantum mechanical part of the electronic energy (the "exchange-correlation" or "XC" energy) by making estimates of the XC energy from the homogeneous electron gas (this is the "local density approximation" or LDA). Kohn and Sham also showed that for practical implementation (connected with difficulties in estimating the kinetic energy of the electrons), it was necessary to retreat from a formulation based soley upon the density, and introduce single-particle orbitals "Kohn-Sham orbitals" into the problem, such that the sum of the squares of these orbitals yielded the electronic charge density. The mathematical structure of the resulting "Kohn-Sham" equation is that of the self-consistent field Hartree equations with a complicated potential, which can only be obtained approximately. Unlike the conventional Hartree approximation, the theory includes robust information about the exchange and correlation contributions to the ground state through the LDA. In simplest terms, what DFT in the LDA enables is a remarkably good mapping of the interacting ground state many-electron problem onto an effective one-particle problem. This is an enormous simplification, and makes the calculation of total energies, forces and charge densities practical. We have published elementary outlines of DFT elsewhere[4]; for a current and authoritative treatment see the new book by R. M. Martin[5].

## Forces from Electronic Structure.

In the seventies, a few researchers began to take density functional calculations seriously to compute properties of materials. An important step toward the idea of density functional simulation of complex materials was the calculation of Kunc and Martin[6], who used DFT to compute phonon dispersion. The fact that the method worked at a predictive level with no experimental inputs was a harbinger of the developments of the next two decades. In 1985, Car and Parrinello[7] published a celebrated paper which gave the first "recipe" for combining density functional theory with practical dynamical simulation. Contemporary with this, Sankey and Allen were developing local basis methods for dynamical simulation, initially using an empirical tight-binding Hamiltonian[8], but also with density functional theory using atomic like orbitals as basis functions[9]. Both approaches exploit the key idea of *obtaining the interatomic forces directly from the electronic structure of the material.*

Highly optimized plane wave basis implementations of DFT are available, and are the "gold standard" for converged LDA calculations[10]. They are computationally expensive for many problems, especially for atoms with compact atomic orbitals requiring "hard" pseudopotentials [5] (like oxygen, carbon and transition metals). The plane wave representation also has the limitation that there is no practical way to adopt the so-called "order N" methods (which have memory and CPU demand scaling linearly with the number of atoms N). For benchmark or other high precision calculations, however, there is no substitute for a good plane wave code.

For modeling large systems (for our purposes, involving more than ~150 atoms[11]), a necessity in the field of disordered systems, we find that sophisticated local basis methods are ideal. Two such codes are SIESTA[12] and FIREBALL[9]. Both codes employ numerical local orbitals (essentially atomic-like orbitals) and rich basis sets (the former including "polarization" functions which are unoccupied in the atomic ground state, but which can be necessary to represent Kohn-Sham orbitals in complex environments). These codes can be crudely thought of as "*ab initio* tight-binding" approach to electronic structure. The delicate point for local basis codes is ensuring that the basis is "complete enough" *for the topology being explored*. For example, crystalline Si and the vast majority of sites in a realistic model of amorphous Si are well described by a minimal (one s and three p orbitals per site) basis. On the other hand, to describe defects properly one often needs much richer basis sets. In general, strange topologies, and reduced topological or chemical order require more delicate (and unfortunately more expensive) calculations[13]; and this is a harder issue to test with local orbitals than plane waves, where completeness has one 'knob': the plane wave cutoff. In special circumstances (when there is a large optical gap free of defect states) it is possible (although *not* trivial) to use the "order N" methods with local basis codes, which can make very large and difficult problems possible to tackle.

It is difficult to exaggerate the value of accurate simulation of materials. It is one of the most important developments in theoretical/computational physics in the last 30 years. It has enabled deep insights into the structure of biomolecules, surfaces of every imaginable kind, the character of defects, and of course the properties of complex materials like amorphous solids. Given a particular structure, it becomes possible to infer local details of the electronic character, phonons, optical properties, defects etc. The information that is obtained far exceeds what experiment can provide, because local information is automatically available; we are not limited to spatial average results, but can learn the electronic structure or vibrational excitations at a given site (for example). In conjunction with careful comparisons to experiment (to verify that the structure is consistent with measured properties), a powerful new approach to condensed matter physics has emerged. The award of the 1998 Nobel Prize in chemistry to Walter Kohn and John Pople reflects this[14]. Still, it is fair to reflect briefly upon what the methods *cannot* accomplish. The gravest shortcoming is the short time scale accessible to the methods (at the very best nanoseconds, and more realistically picoseconds). This means that it is *not* possible to directly (realistically) simulate a process like quenching from the melt (which in principle involves macroscopic numbers of atoms and may require time scales *qualitatively* longer than what the simulation can provide). An interesting feature of the simulation paradigm is its analogy to experiment. Vast amounts of raw data (atomic positions, velocities, electronic structure, density matrix, eigenvectors etc) are the output at each time step, but one must learn how to extract physically informative data from the plethora of information.

## Structural Modeling

We begin this section with a sermon on modeling. The first job of a theorist modeling any material is to create a reasonable structural model. Such model building is an example of what is sometimes called an 'inverse problem': given some incomplete (experimental) information about a material, <u>infer</u> its structure. Thus, one can begin with measured pair-correlation functions (or static structure factors) and try to find atomic coordinates reproducing the measurements. The information in the pair distribution function of amorphous materials is remote from providing information adequate to uniquely specify coordinates in a model, as we discuss in connection with the "Reverse Monte Carlo" technique below. This then begs the question: "What collection of experiments does uniquely specify a model of an amorphous material"? The answer is almost certainly that no currently conceivable set of experiments leads to a unique model. It may be the case that a set of experiments by themselves might *usefully* constrain the coordinates to a "representative" subspace of all possible models. Probably other external information (like chemical bonding constraints) is required to form a realistic model. For materials in which even the chemical ordering is murky, it is likely that accurate first principles modeling will be required for model building.

Because most measured quantities are averaged over large numbers of atoms, each with a unique environment, the outcome of experiments is usually smooth and rather featureless, and therefore carries limited information. The contrast is to a field like protein crystallography, in which exquisite detail is provided by diffraction data with a palisade of near delta functions. This information leads to impressive reconstructions of large *crystal* unit cells. Information theory[15] may be used to gauge the "information content" of such data through the use of the information entropy[16]. The smooth curves obtained from amorphous materials carry much less information and therefore specificity about microstructure. This is rather ironic, since this is the type of material for which structural information is most desperately needed.

Another fundamental limitation of the conventional diffraction measurement is that it is sensitive only to pair correlations. The one promising exception to this rule is the so-called Fluctuation Electron Microscopy due to Treacy, Gibson and Voyles[17]. One should view all such diffraction experiments on amorphous materials as providing "sum rules" which must be satisfied, but are inadequate by themselves to identify a model. One important case in which experiments might turn up highly specific information is from spectroscopy (electronic, magnetic or optical). For example, electronic defect states in the optical gap are often very localized and at well defined energies (which may imply very specific defect conformations). Such information is invaluable because of the smoothness (and associated microscopic non-specificity) of the other experiments. Potentially a scanning surface probe like STM may also yield local structural information.

So, while we argue that experiments may not adequately constrain atomic positions, we also must emphasize that any model which is to be believed must reproduce all the experiments available. Obvious as this is, a many papers celebrate agreement with one incomplete measure (a single experiment) and ignore other experiments. In fairness, it may be difficult to match all of this information, but it is a goal we must strive for.

## Cook and Quench Method

The intuitively natural way to make a computer model of an amorphous or glassy material is to *pretend* to mimic nature: form an equilibrated liquid, "cool" it through the glass transition (if there is one), then relax the arrested network so that the forces vanish on all the atoms. We name this with some irreverence "cook and quench". As discussed above, there is no reason to believe that this process is to be taken literally as a simulation of laboratory glass formation because the time scales are completely different—the simulation cannot explore a similar volume of configuration space as that visited by experiment. In fairness, one should also point out that our formulation exaggerates the defects of cook and quench, since some of the long time scales encountered in experiments arise from the finite thermal conductivity of the liquid (so that it takes atoms far away from the ice bath a "long time" to learn that they and the other atoms in the liquid have been tossed into an ice bucket, and behave accordingly). In a simulation by contrast, the "quenching " is done uniformly throughout the sample and involves all the atoms cooling down together. Still, there is a fundamental difference between simulation and experiment!

Empirically, there have been successes for cook and quench. Most notably this has been observed for silica and certain chalcogenide glasses. Advocates of the method assert that it is "unbiased"(not forcing the system toward any *a priori* preferred result). This is not entirely true, since the method is clearly biased to incorporate too much liquid character into the solid state. The method has failed for amorphous silicon and complex (ternary) chalcogenide glasses[18]. It is reasonable to conjecture that cook and quench should work when (1) the structure of the liquid is essentially similar to the structure of the glass, (meaning that similar fundamental units or "building blocks" are present in both) and (2) the ordering is quite local (which amounts to saying that the building blocks from which the glass (and liquid) is composed are quite small). The failure of cook and quench to produce reasonable models (that is, with a small concentration of coordination defects) of Si is probably connected to the fact that the liquid is ~6-fold coordinated and a metal[19], whereas the amorphous phase is a tetrahedral insulator with a concentration of non four-fold atoms less than 0.01% (even *sans* hydrogen). In g-GeSe$_2$, a classic stoichiometric chalcogenide glass, we have seen that cook and quench leaves signatures of excessive liquid-like character in the static structure factor. For large q, S(q) for the cook and quench models decays away too rapidly relative to experiment. We discuss the performance of cook and quench further below in conjuction with other modeling schemes.

## Wooten-Weaire-Winer Methods

For the peculiar case of tetrahedral amorphous insulators a-Si and a-Ge, there is no doubt that the finest models are made with the "WWW" technique, due to Wooten, Winer, and Weaire [20]. This technique is essentially a Monte Carlo modeling approach with very specific rules for Monte Carlo moves. In the original version, one starts with a perfect diamond structure, and then adopts the "WWW bond transposition" or bond switch. For a bonded pair of atoms BC a pair of nearest neighbors A and D is chosen, so that A is the neighbor of B and not the neighbor of C, and D is the neighbor of C and not the neighbor of B. Then bonds AB and CD are broken (deleted from the bond lists for atoms B and C) and new bonds AC and BD are created (added to the appropriate bond lists), i.e. atoms B and C exchange neighbors. This procedure effectively introduces five- and sevenfold rings—which are a characteristic structural feature of the CRN—in the network *while preserving the four-fold coordination.*

Monte Carlo moves are accepted in Metropolis fashion with Keating springs as the interatomic potential. In practice, the method is not trivial to implement, as one needs to introduce "sufficient" disorder (so that the system does not return to a crystalline state) and a proper simulated annealing scheme to produce an optimal network. Recently, Mousseau and Barkema [21] have shown that it is not necessary to start with diamond – a completely random configuration leads ultimately to topologically identical networks as those obtained from the randomized crystal. Carefully devised WWW networks are in remarkable agreement with experiment on structure, electronic structure and dynamics. Our belief is that the method is successful for two reasons: 1) the moves identified by WWW are in fact quite physical (as shown by the "Activation Relaxation Technique" (ART), discussed later) and 2) the method compels the system to retain four coordination, and indeed to force bond angles close to the tetrahedral angle (through the bond angle "springs" in the potential) . This second condition amounts to constraining the optimization of the network to satisfy *a priori* information (which can be inferred from optical and other measurements). It is notable that a very naive potential is adequate when the simulation is performed with suitable *a priori* information (especially because cook and quench with *ab initio* interactions is not very successful for silicon). The method may also be improved to model heterogeneous materials with crystalline regions[22] (these appear to be important for photovoltaic applications).

## Decorate and Relax

In this section, we present an approach for modeling binary glasses beginning with models of tetrahedral amorphous semiconductors and report new models of glassy $GeSe_2$, $SiSe_2$, and $SiO_2$. The topologies of our models are analyzed through partial pair correlations and static structure factors. Our approach is extremely simple and faster than traditional cook and quench simulations and emphasizes the importance of correct topology of starting structure for successful modeling.

For some of the simulations reported in this section, we used FIREBALL, a density functional code in the LDA developed by Sankey and coworkers. The basis set is minimal (for these systems, one *s* and three *p* slightly excited pseudoatomic orbitals per site or "single zeta" in the language of quantum chemistry). In its original form only weakly ionic systems could be treated; self-consistent versions have been developed recently [23]. These approximations perform exceptionally well for chalcogenide systems. The other code is SIESTA [24], which has broad flexibility with respect to basis set, density functional, and simulation regime. We employed SIESTA for silica because the extreme ionicity of the material, and also to easily check the importance of density functional, basis set and spin polarization. In the end, we found that relatively simple approximations (self-consistent LDA and a single zeta basis) were adequate. Even using soft pseudopotentials, we found that a 150 Ry cutoff was needed for evaluation of the multicenter matrix elements.

We made models of $GeSe_2$, $SiSe_2$ and $SiO_2$ glasses by starting with a defect-free (fourfold coordinated) 64-atom supercell model of *a*-Ge made with the WWW method [25,26]. Characteristic of an amorphous column IV material, this model has bond angles tightly centered on the tetrahedral angle, and has a topology presumably unrelated to g-$GeSe_2$, g-$SiSe_2$ and g-$SiO_2$. We decorated all the IV-IV bonds with a bond-center VI, and rescaled the coordinates to the experimental density of g-$GeSe_2$, g-$SiSe_2$ and g-$SiO_2$ respectively. The 192 atom models of g-$GeSe_2$ and g-$SiSe_2$ were then quenched with FIREBALL to the nearest minimum. The 192-atom model of g-$SiO_2$ was relaxed with SIESTA. We name this scheme "decorate and relax". The resulting models are in some ways superior to the best models in existence, are remarkably easy to generate, and preliminary work with Chubynsky and Thorpe suggests that the approach may be extended to off-stoichiometric compositions. Such networks have been introduced and explored by Chubynsky and Thorpe[27] to study the vibrational excitations of chemically ordered networks. Vink and Barkema have also explored some related methods in silica [28]. We then extended the calculations to decorated models with 648 atoms (starting with 216 atom WWW cells).

The decorated models have general similarities and origins that we illustrate for the case of $GeSe_2$. In Fig. 1 we report the static structure factor for decorated diamond (simply Ge on a diamond lattice with bond center Se added, without relaxation and rescaled to the experimental density of glassy $GeSe_2$) and WWW a-Ge similarly decorated and rescaled. In both models we note the obvious presence of a strong, sharp prepeak in S(Q). In the crystalline decorated diamond model, the so-called first sharp diffraction peak (FSDP) arises from the <111> Bragg peak of the structure. This prepeak is very similar to the prominent

FSDP feature of glasses. The existence of this peak in both models shows that our starting models already exhibit the intermediate range order associated with the FSDP.

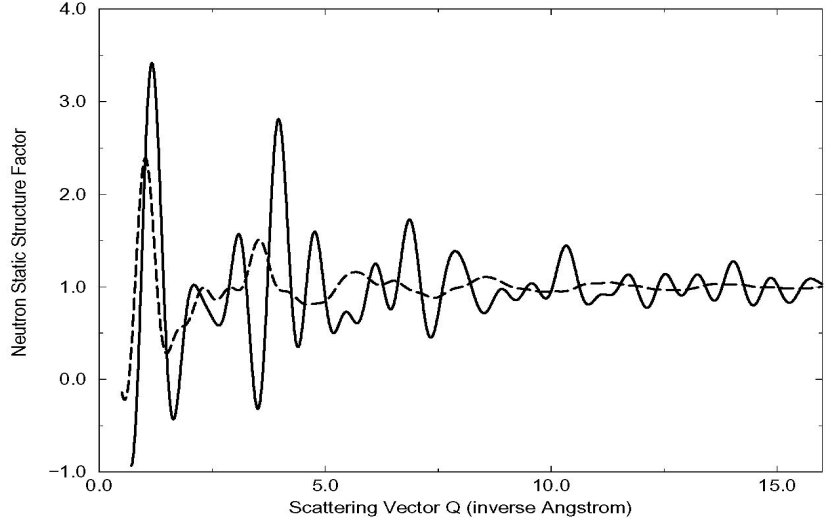

**Fig.1. Calculated total neutron structure factor S(Q) of unrelaxed "decorated" glassy GeSe$_2$ (dashed line) and unrelaxed "decorated" diamond GeSe$_2$ (solid line). We used scattering lengths of b$_{Ge}$=8.185 and b$_{Se}$=7.970 fm.**

The structure of these models have been analyzed by computing the partial Faber-Ziman structure factors. In earlier work, we compared the results for the Faber-Ziman structure factors S(q) vs. experiment [29,30] the earlier model of g-GeSe$_2$ [31,32] and the new model. The decorated model is at least as good as the previous model and comparable to the models of Massobrio and co-workers [33,34]. While the new model has strong similarities manifested in the partial structure factors, and essentially similar topological/chemical ordering to the cook and quench model, a key difference of the "decorated" model is the persistence of correlations in S(q) beyond 10 Å$^{-1}$ in unique and pleasing agreement with experiment, whereas the earlier model displays a more rapid decaying amplitude for large q (see Fig. 2). Our interpretation is that the cook and quench model was too "liquid like" - precisely the kind of artifact one might expect from rapidly quenching a liquid on the computer. The new model has 86% heteropolar bonds, with the homopolar bonds Se-Se (13.5%), except for a single bond (0.5%). Ge was 78% fourfold, 19% threefold and 3% twofold, numbers quite consistent with the earlier model. The agreement of such details with the earlier model seems remarkable, as the simulation processes were so very different. We also computed the vibrational and electronic states densities (EDOS) and found them to be very similar to the earlier model of Cobb[32]. We have reported the details on GeSe$_2$ elsewhere[35].

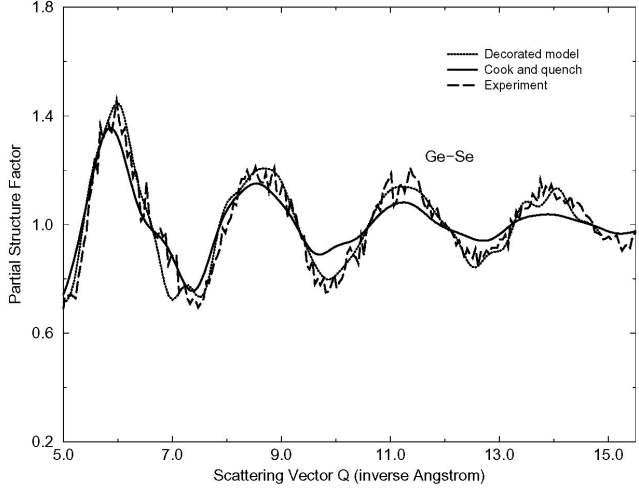

**Fig. 2. A blow-up of the neutron Ge-Se partial structure factor for 192 atom model (Ref. 29) of glassy GeSe$_2$. Note the rapid decay of S(Q) for the quenched model and improved agreement with experiment (Ref. 31,32) for the decorated WWW model.**

'Decorate and Relax" was also used to generate a model of g-SiO$_2$. The properties of our model have been studied through the neutron static structure factor and the partial pair-correlation function. In Fig. 3 we illustrate the real-space partial pair correlation function and compare to experiment[36]. The position of the first peak in $g_{SiO}(r)$ gives the Si-O bond length to be 1.62±0.02 Å. The corresponding experimental value from neutron-diffraction data[36] is 1.61±0.05 Å. The nearest-neighbor O-O distance from Fig. 3 is 2.65±0.05 Å and the corresponding experimental value, inferred from neutron-diffraction data, is 2.632±0.089 Å. By taking the Fourier transform of the pair-correlation function we compute the neutron static structure factor $S_N(Q)$ which may be directly compared to its experimental counterpart[37]. The total static structure factor of our model, together with the one obtained by neutron diffraction experiments[36] is presented in Fig. 4. Our results are again in reasonable agreement with experiment. The position of the FSDP of our new "decorated" model coincides with the experimental one. The system has no homopolar bonds, as one would expect from the chemistry of silica. We also use decorate and relax to produce a 648-atom model of SiO$_2$ starting with a defect free 216-atom model of a-Si. The static structure factor of this model along with experiment and the 192-atom model is plotted in Fig. 4, in essentially perfect agreement with experiment. The discrepancy between the 192 and the 648-atom models arises from finite size effects, since the same Hamiltonian and procedure was used to generate both models. It is of some interest that the only substantial difference between the 192 and 648 atom models was near 2.0 Å$^{-1}$ in the minimum after the first diffraction peak. The only notable remaining discrepancy between theory and experiment appears near 12 Å$^{-1}$, and is similar for both models (and so is not due to a finite-size effect). The electronic density of states shows a state-free optical gap consistent with silica and the usual limitations of the LDA for predicting the gap.

We extended the method to g-SiSe$_2$. The properties of our "decorated" model are studied through the neutron static structure factor and the partial pair-correlation function. In Fig. 3 we present the real-space partials pair correlation function of our model. There is good agreement between our simulated results and the earlier model[38]. The sharp peak in the pair correlation function $g_{SiSe}(r)$ is due to the largely predominant heteropolar Si-Se bonding. In $g_{SiSi}(r)$ the peak at 2.4±0.05 Å is due to Si-Si homopolar bonds. The main peak in the $g_{SeSe}(r)$ stems from the intratetrahedral second neighbors Se-Se distances while the small peak at 2.4 Å is indicative of homopolar Se-Se bonding. The calculated neutron scattering structure factor (Fig. 5) shows very good agreement with experiment[39]. Analysis of S(q) reveals close agreement once again between experiment and theory for large q, as seen in GeSe$_2$. The optical gap is consistent with experiment.

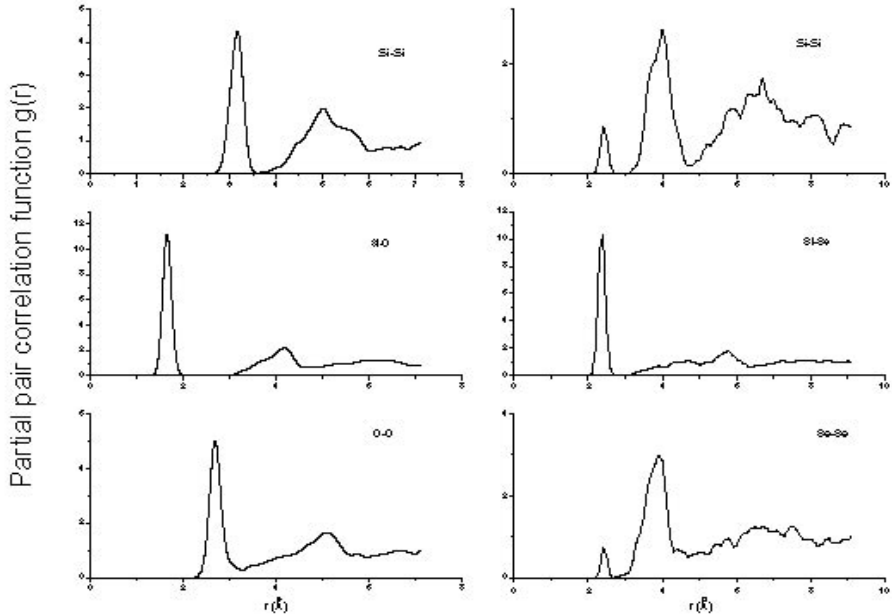

**Fig. 3: Partial pair distribution functions** $g(r)$ **vs. r in** $g-SiSe_2$ **(left panel) and in** $g-SiO_2$ **(right panel).**

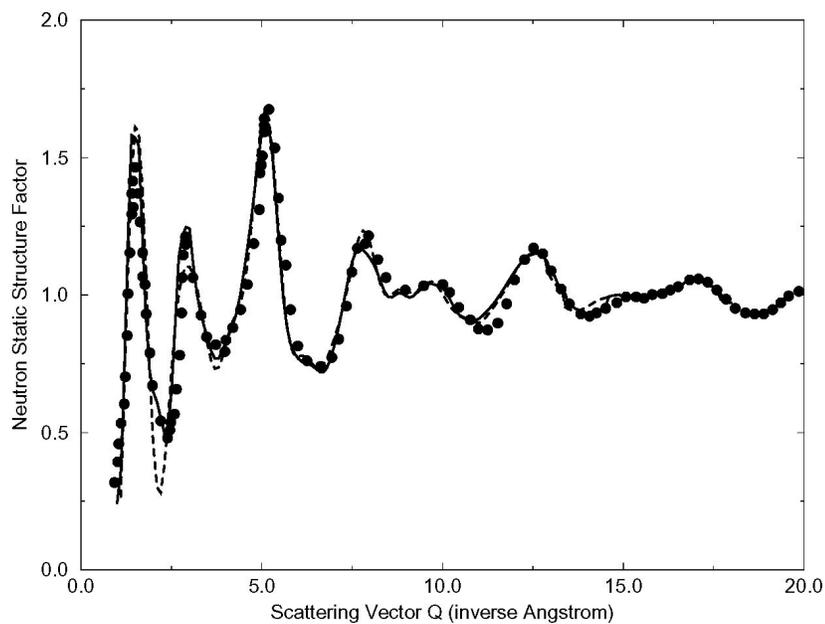

**Fig. 4. Calculated total neutron static structure factor S(Q) of glassy $SiO_2$ (dashed lines are for 192-atom model and solid lines are for 648-atom model) compared to experimental data (Ref. 36) (filled circles). We used scattering lengths of $b_{Si}$=4.149 and $b_O$=5.803 fm. Note the close agreement between experiment and the 648 atom model.**

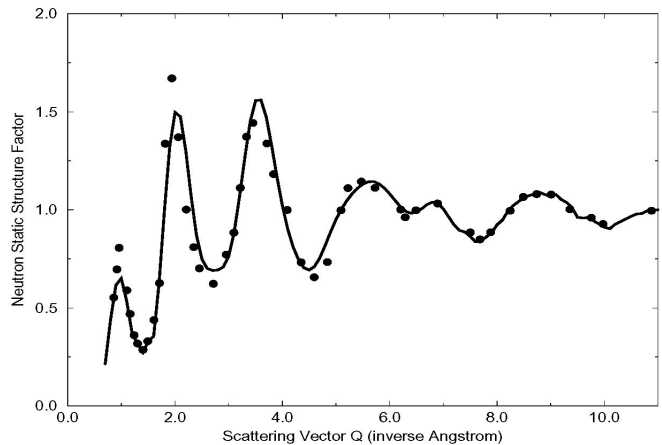

**Fig. 5. Calculated total neutron static structure factor S(Q) of glassy $SiSe_2$ and experiment (Ref. 39).**

The key point concerning Decorate and Relax is that with a simple idea we were able to generate models of IV-VI$_2$ glasses close to or improving upon the best "cook and quench" models, with the appealing feature of a proper asymptotic behavior for large Q (close agreement between experiment and theory was seen for large q for all of the models we fabricated with the method). This decay of correlations for large q is one signature of the state of the material: for crystals the correlations persist to very large q, for liquid they decay very rapidly and the glassy state is somehow intermediate. For the three systems we have studied, it appears that Decorate and Relax picks this special decay up reliably, whereas cook and quench does not. The method spares us from expensive calculations of cook and quench technique (many phases with many time steps each). Overall the decorated scheme is much faster than the traditional methods (at least 10 times faster for a given interatomic interaction). To the extent that no scheme can be claimed to mimic the *physical* process of glass formation, this method should be evaluated by its success in reproducing the known experimental information. We have seen that the scheme produces possibly the best model of silica to date involving a reasonable number of atoms (648) and accurate forces (from SIESTA).

## Activation-Relaxation Technique (ART)

ART[21,40] is designed to overcome the limitations of MD, especially the profound limitation of time scales. ART deliberately seeks saddle points in the energy landscape and moves configurations in a concerted way over such saddle points to new basins (and therefore potentially to a new local minimum). The method has been applied to many problems, particularly to structure of a-Si, where it was possible to determine the dynamical events on time scales vastly exceeding those of conventional MD. It is quite remarkable that ART "discovered" the WWW moves as dominant in a-Si, and this is a strong hint at why WWW is so successful. Maybe the way to generalize WWW to other systems (really, how to generate the right moves for other systems) is to use ART with accurate interatomic potentials (probably *ab initio*) to identify the dominant moves, then use the WWW simulated annealing with these moves and a simple potential. Probably other information (like correct chemical ordering or coordination) also would need to be built in.

## "Reverse" Monte Carlo Methods

An appealing conceptual basis for model building is to form an atomistic model that agrees with all acceptable experiments. In one's imagination, this corresponds to filling a large box with atoms, and moving them in some way until the coordinates reproduce all the known experiments. This is essentially the idea underlying the so-called 'reverse' Monte Carlo (RMC) approach[41,42]. Unlike most of the traditional methods that we have described so far, the most striking feature of RMC is its' ability to model disordered materials without the need of any potentials. At its simplest, RMC is a technique for generating structural configurations based on experimental data coupled with a set of appropriately chosen constraints. Originally developed by McGreevy & Pusztai[43] for modeling liquid and glassy materials, the method has been developed in recent years to model crystalline systems as well[44]. Here one starts with a suitable configuration, which may or may not be completely random, and prescribes a set of rules for evolution of the atomic positions. The atoms are displaced randomly using the periodic boundary condition until the input experimental data (either the structure factor or the radial distribution function) match with the data obtained from the generated configuration. This is achieved by minimizing a cost function which consists of either structure factor or radial distribution function along with some appropriately chosen constraints to restrict the search space. Consider a system having N number of atoms with periodic boundary condition. One can construct a generalized cost function for an arbitrary configuration by writing :

$$\xi = \sum_{j=1}^{K} \sum_{i=1}^{M} \eta_i^j \left[ F_E^j(Q_i) - F_c^j(Q_i) \right]^2 + \sum_{l=1}^{L} \lambda_l P_l$$

Here, both terms are positive definite, and the first term vanishes when a model exactly reproduces all experiments considered and the second term is zero when other constraints (such as a particular topology or chemical order) is exactly satisfied. More specifically, $\eta_i^j$ is related to the uncertainty associated with the experimental data points as well as the relative weight factor for each set of different experimental data. The quantity $Q$ is the appropriate generalized variable associated with experimental data $F(Q)$ and $P_l$ is the non-negative penalty function associated with each constraint. In order to avoid the atoms from coming too close to each other, a certain cut-off distance is also imposed which is typically of the order of interatomic spacing and is usually obtained from the radial distribution function. This is equivalent to adding a hard sphere potential cut-off in the system which prevents the catastrophic building up of potential energy. If some of the characteristic features of the structure (e.g., bond angle, coordination number etc.) are available from experiments, one can implement a Monte Carlo scheme by incorporating these constraints to obtained a final structure. Although RMC has been applied to many different types of systems – liquid, glasses, polymer and magnetic materials, questions are often raised about the reliability of results obtained from RMC simulation[43]. The method has never been accepted without some degree of controversy and the most popular criticism is the lack of unique solution from RMC. The practical utility of the structure obtained from RMC simulation usually suffers from two major drawbacks. First of all, RMC can produce multiple highly distinct configurations having the same pair correlation function. This problem is acute when only radial distribution or structure factor is used to obtain the configuration with no or a few constraints. In this case, the 'energy' landscape defined by the cost function in the equation above deviates significantly from the true energy surface and the system evolves only to converge to a fictitious local minimum. The second problem, which is related to the first, is that there is no guarantee that the generated structure corresponds to a local minimum on the multidimensional true energy landscape. This is not surprising in view of the fact that only pair correlation function or structure factor is used in modeling the structure while there exists an infinite hierarchy of higher order correlation functions which are not directly connected with experimental data. Therefore, in absence of sufficient information, RMC can only produce the most disordered structure consistent with a given set of experimental data. This can be illustrated by taking amorphous silicon as an example. The key information that is necessary to model amorphous silicon is that the bond angles among the neighboring atoms as well as their coordination number must be strongly constrained. By themselves, the structure factor or the radial distribution function can not provide such information and consequently RMC can not generate configurations having *only* the topology of amorphous silicon but a mixture of all those that are consistent with structure factor or radial distribution function used in the simulation. Unconstrained RMC is the ideal tool to infer how much information is available in a given experiment.

Recent work on reverse Monte Carlo simulation[45] by the present authors have indicated that most of the shortcomings mentioned above can be eliminated by constructing a suitable cost function consisting of a set of penalty functions. As we have discussed in the preceding paragraph, the success of the RMC modeling lies in the identification of suitable constraints and the efficiency with which information contained in the constraints can be implemented via Monte carlo scheme. It is to be noted that while there is no limit to the number of constraints that can be included in the system, there is no guarantee that mere inclusion of more constraints will necessarily give better results. Forcing a completely random configuration with too many competing constraints may cause the configuration to be trapped in the local minimum of the function $\xi$ and may prevent the system from exploring a large part of the search space. This is a version of the constrained optimization or "traveling salesman" problem, a persistently vexing puzzle of applied mathematics. For amorphous silicon, an optimally constrained cost function taking into account the chemical and geometrical nature of the bonding can be written as :

$$\xi = \sum_{i=1}^{M} \lambda_1 [F_E(x_i) - F_c(x_i)]^2 + \lambda_2(t)\,(\theta_0 - \theta(t))^2$$

$$+ \lambda_3(t)\,(\delta\theta_0 - \delta\theta(t))^2 + \lambda_4 \left[1 - \Theta(x - x_c)\right] + \lambda_5 [\phi_0 - \phi(t)]^2$$

**where,**

$$\theta(t) = \frac{1}{N_\theta(t)} \sum_{i\{j,k\}} \theta_{ijk} \quad \text{and} \quad \delta\theta(t) = \sqrt{\langle (\theta - \theta_o)^2 \rangle}$$

In the above equation, $\theta(t)$ and $\delta\theta(t)$ are the average angle and the rms deviation at time t (here, 'time' refers only to the progress of the Monte Carlo minimization ) while $\phi(t)$ and $\phi_o$ are the instantaneous and proposed concentration of the 4-fold coordinated atoms. It is important to note that each of the terms in the equation above should decrease ideally to zero during the course of simulation.

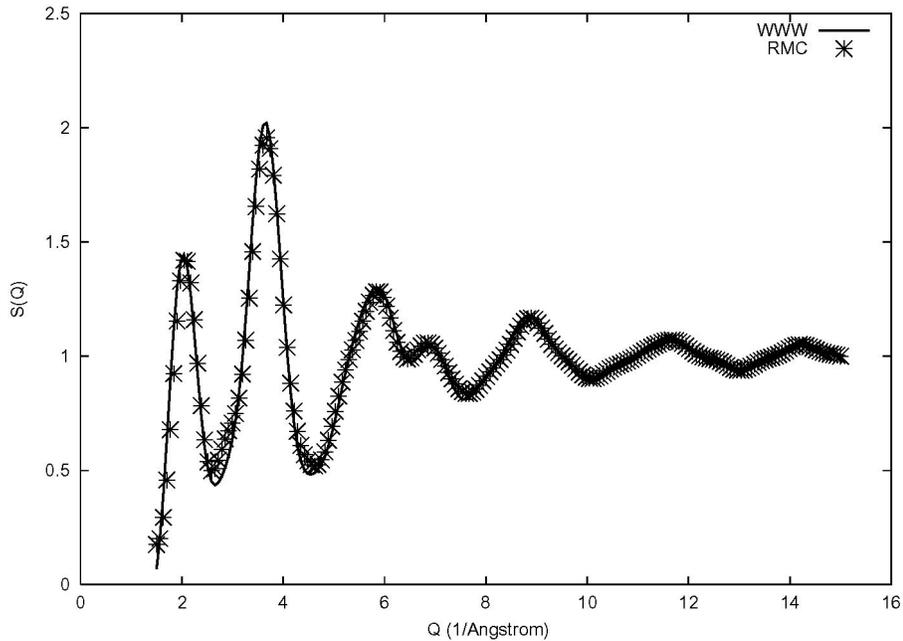

**Fig. 6. Structure factor obtained from a fully constrained model of RMC. The solid line is the structure factor for a WWW model (which is essentially identical to experiment).**

We have implemented RMC with a completely random starting configuration. This eliminates any possible local ordering or *a priori* bias that might exist in the starting structure (e.g., randomized diamond structure may retain the memory of tetrahedral ordering). Based on experimental consideration, the model includes only the key features of amorphous tetrahedral semiconductors – an average bond angle of $109.5°$ having a rms deviation of $10°$ observed experimentally. For 216-atom model, the size of the box is 16.008Å, which corresponds to number density 0.0526 atom/Å$^3$. The initial configuration is generated randomly so that no two atoms can come closer to 2.0Å. The configuration is then relaxed by moving the atoms to minimize the cost function

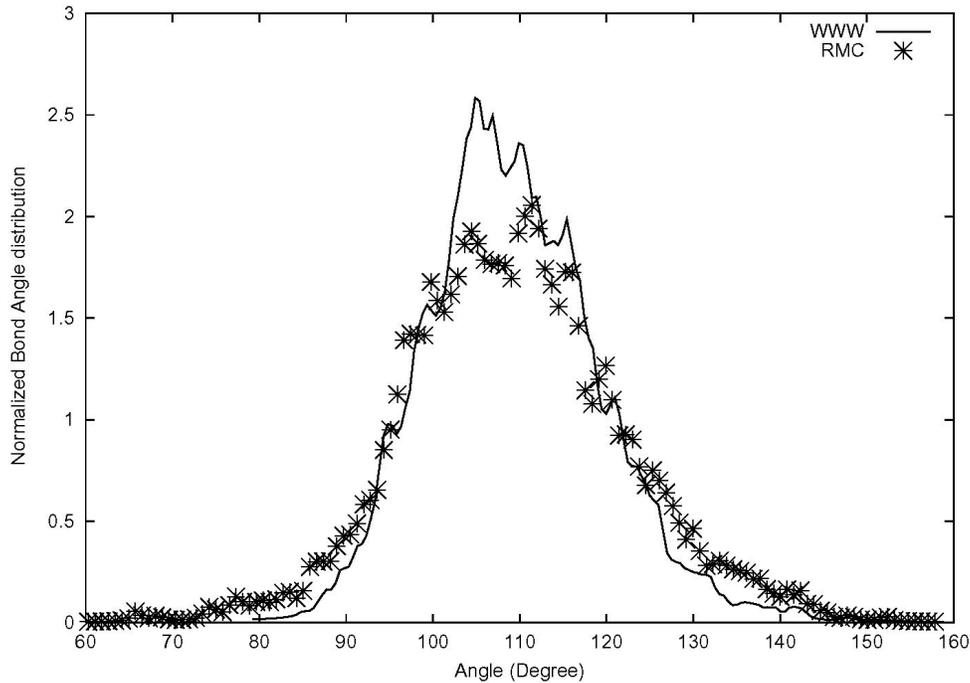

**Fig. 7. The bond angle distribution functions for *a*-Si from constrained RMC (points) and WWW model (solid line). The solid line is obtained from a WWW model having rms deviation of $9.5°$. The rms deviation for the constrained RMC model is found to be $12.3°$.**

The results for the model including all the constraints are presented in Figs. 6-8. We have plotted the structure factors for the constrained RMC and WWW model in Fig. 1. The agreement is excellent both at small and at large value of $Q$. It is clear from the Fig. 7 that the distribution obtained from the RMC model follows the tetrahedral character observed in amorphous semiconductors. The average bond angle in this case is found to be $109.3°$ with rms deviation of $12.35°$. An important aspect of the bond angle distribution in Fig. 2 is that most of the angles are lying between $70° - 150°$ compared to $80° - 140°$ in WWW case (solid line). We have plotted the electronic density of states (EDOS) for the constrained model in Fig. 8. The EDOS appears with all the characteristic features of *a*-Si with the exception of a clean gap in the spectrum. This behavior is not unexpected in view of the fact that 88% of the total atoms are found to be 4-fold coordinated with an average coordination number 3.84. The presence of the defect states makes the gap noisy and at the same time the use of LDA underestimates the size of the gap. It is, however, curious to note that the average coordination number obtained from our model is close to the experimental value of 3.88 reported by Laaziri *et al.* [46].

What our calculation reveals is that with patience (identifying the right constraints and extended minimization), RMC may be used as a tool to make useful new models of amorphous materials. We believe that it also has utility as a way for making simple models as inputs to first principles calculations with their restricted time scales.

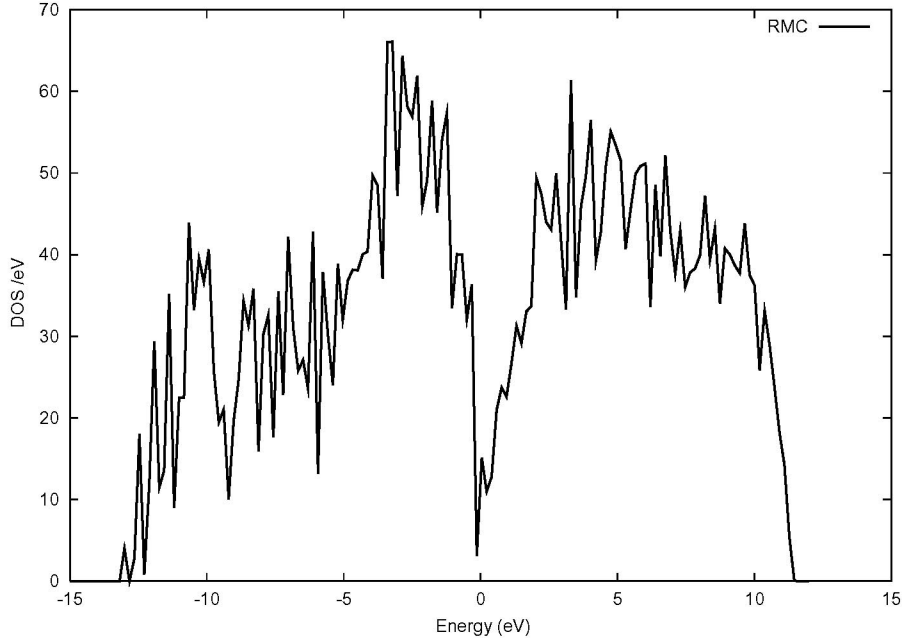

**Fig. 8. The electronic density of states of *a*-Si obtained from constrained RMC simulation described in the text. SIESTA was used to model the electronic structure. The Fermi level is at E=0.**

## Electronic Properties

The dynamical behavior of electrons in disordered materials has been studied at least since the seminal work of Anderson on localization. There is now a deep understanding of the zero-temperature Anderson problem in all dimensions. For *finite temperature*, far less is certainly known. The reason for this is the difficulty of handling the time dependence in the electronic Hamiltonian arising from lattice vibrations. It is clear from hopping theory that the lattice dynamics can drive electronic diffusion (hopping), and some features of the temperature dependence can be inferred from simple arguments such as Mott's variable range hopping. Understanding of transport in amorphous materials depends upon a microscopic understanding of hopping and electron dynamics. See Mott and Davis[47] for a classic review.

Earlier work has shown that there is always a large electron-phonon coupling for localized or partly localized electron states. This can be seen from calculations of deformation potentials (which measure the response of a given electronic eigenvalue to a lattice distortion induced by phonons). Also, if one performs a standard Born-Oppenheimer thermal MD simulation of a material like a-Si, it is found that the electronic Kohn-Sham eigenvalues fluctuate quite dramatically near the optical gap, where the states may be localized. The root-mean-square (RMS) fluctuation of these energy levels can be several tenths of an *eV* at 300K, an effect which greatly exceeds *kT*.

An alternative approach is to use the time-dependent Schrodinger equation to study the time evolution of an electron packet. The idea is that at finite temperature the lattice is in thermal motion. Thus, in an adiabatic picture, the one-electron (density functional) Hamiltonian of the system is time dependent (just because of the motion of the atoms). Because of this time-dependence, one must use the full time-dependent Schrodinger equation to see the evolution of the electron. Of course this is the essence of thermally driven hopping. If there is an electric field present in the material, it is also the beginning of a non-adiabatic theory of carrier transport beyond linear response.

## Experimental Probes

A key probe of electron dynamics in disordered systems is the time of flight experiment for the drift mobility. Here, a thin film sample is sandwiched between two blocking contacts across which is maintained a potential drop, and a laser flash is used to create carriers, which are swept toward the appropriate electrode. The basic physics of the experiment is that the electrons or holes encounter a variety of traps as they drift through the sample. A typical analysis of data from this experiment is to attempt to solve a difficult inverse problem; extracting the density of states (DOS) from the post-transit current (defined roughly as the current after the number of free carriers has been reduced by 1/2.) The difficulty of this analysis is at least two-fold: (1) temperature dependence is usually modeled crudely, or even neglected, even though it is clear[48] that there is strong thermal modulation of the trap energies, and also this is strongly dependent upon where the state lies relative to the Fermi energy[49] (2) Almost nothing is known about the structure of band-tail states and the consequences to the capture cross-sections. A recent experiment revealed that time of flight measurements were sensitive to the microstructural details originating from different growth conditions. For example, the hole mobilities in expanding thermal plasma deposited a-Si:H is a factor of *10* higher than for conventional plasma-enhanced chemical vapor deposition samples (while no such effect is seen for electron mobilities)[50]. It is argued that this implies the existence of crystalline inclusions. The atomistics of this observation is not well understood. A successful phenomenological framework explaining a number of experiments is that of Overhof and Thomas[51] and Baranovskii and coworkers[52] who emphasize the importance of electron-phonon effects, and have used Monte Carlo methods to simulate band-tail hopping. These approaches offer an ideal example of where additional theoretical information (for example on the energy dependence of the capture cross section and on thermal modulation of electronic energies) can aid the interpretation of transport experiments.

Ultrafast pump-probe laser spectroscopy and spectral hole burning experiments are well known in systems like quantum wells and nanostructures and comprise a standard tool there to determine fast (as fast as 10 fs!) carrier dynamics, study the thermalization of hot carriers, and spectral and spatial diffusion of carriers.

These experiments, and particularly spectral hole burning, (in which a small part of the spectrum is saturated and the resulting "hole" spectrally diffuses into neighboring energy states) are ideal for comparison to our simulations. By tracking the time evolution of what is initially an eigenvector of the Kohn-Sham Hamiltonian from the TDSE, we can observe the spectral and spatial electronic diffusion. Systems with disorder are especially interesting for this type of work – there are localized (mid-gap) states, less localized band tail states (with a mobility edge, at least at T=0). The presence of phonons makes the whole process much richer with thermally modulation of the energy levels, and delocalization processes.

Dexheimer and coworkers[53] have begun ultrafast pump-probe studies on a-Si:H and a-SiGe alloys. The first experiments have used an 800nm (1.55 eV) photoexcitation, which should yield information on the tails, but in this direct approach a two-photon (3.1eV) process dominates, which puts carriers well into the extended states. A very fast carrier thermalization is assigned to phonon emission on a 150 fs time scale and a relaxation on a 230 fs time scale is attributed to phonon equlibration. Interestingly, the phonon emission process seems to be *faster* in the amorphous material than crystalline Si. Such times scales (and significantly longer ones) are easily accessible to our calculations. It is conjectured that a carrier-carrier scattering time occurs on even faster time scales, currently unresolved by the experiment. This process if of great interest, and we will model it with our TDSE with excited carrier-ion coupling. It is of special interest to probe the gap and tail states, since these are of key importance to transport and optical properties, but this is not entirely straightforward experimentally (multi-photon processes dominate over the small tail and gap density of states). Nevertheless, preliminary experiments probing the far infrared (a few *meV*), may provide direct information about carrier trapping and dynamics near the mobility edge. To aid in the interpretation of these measurements, a good estimate is needed for the optical absorption cross section for the gap, tail and extended states. For us, this amounts to computing dipole matrix elements of initial and final states from our simulation.

## Previous Simulations

There have been few attempts to explicitly and realistically model the dynamics of electrons in disordered materials. Two pioneering studies deserve special mention. Selloni *et al*[54] considered mixed classical-quantum dynamics of an excess electron in molten K-KCl by integrating the time-dependent Schrödinger (Kohn-Sham) equation, and solved the coupled equations within a density functional framework with plane wave basis, and obtained rather good estimates for the electron diffusion rates. Nakano *et al*,[55] used a method much like Selloni *et al*, but with simplified interactions between Si atoms (Stillinger-Weber[56]) to study electron diffusion in a model of a-Si. Good agreement with time-of-flight experiments[57] was obtained for electron mobility. The essence of both calculations was the use of the *time-dependent* Schrödinger equation to track the time-development of an initial packet or state in the presence of thermal motion of the ions. There has been recent activity for improved schemes to efficiently integrate the TDSE[58].

## A Density Functional Approach[59]

Electron time evolution in complex materials can now be fruitfully addressed thanks to (1) the availability of reliable density functional methods (we use FIREBALL[9,60]; these methods jointly provide efficient and very accurate descriptions of structure and Kohn-Sham orbitals in complex environments. (2) highly realistic models for amorphous materials (especially Si, $GeSe_2$, $As_2Se_3$ and Se). Here, "highly realistic" implies agreement with structural, electronic and vibrational measurements.

Like others working in this field, we use Kohn-Sham orbitals, which have been shown to be very similar to quasiparticle states from "GW" calculations [61] from Louie's group (such GW calculations provide self-energy corrections to density functional theory in the LDA). For Si, C and LiCl, Hybertsen and Louie[62] found 99.9% overlap between GW states and the Kohn-Sham orbitals. More recently, this conclusion has been substantiated for the Si(001)-2x1 surface and also for a GW study of the metal-insulator transition in BCC hydrogen[63]. We observed a similar state of affairs in our GW calculations of quasiparticle states in Si clusters[64]. Naturally the gap energy is incorrect (too small, usually severely so) for well-understood reasons, but this can be repaired in our systems with "scissors" approximations. Another important finding of this work is that localized states are artificially "fattened" by LDA, and the use of a generalized gradient (GGA) functional leads to a significant increase in inverse participation ratio[65] and accuracy (we have noted a similar difference between LDA and LSDA for localized states in a-Si[66]). On an empirical level for amorphous materials, there are many indications that it is profitable to interpret the Kohn-Sham orbitals "literally" for comparisons to experiments (for example, x-ray photoemission spectroscopy (XPS) measurements on chalcogenide glasses, thermal effects in band tails of a-Si[49], and exciton trapping[67]). We are very interested in developments using "exact exchange" [61] and the GW approximation, which appear to provide a framework for improving upon the use of KS orbitals.

We have tracked the changes in Kohn-Sham orbitals (electronic eigenstates associated with the instantaneous ionic conformation of the system) arising from thermal motions of the atoms computed in the Born-Oppenheimer approximation[49,68]. We have shown that for localized states in the gap or band tails in a-Si, g-$GeSe_2$ and other materials, there is a *very* large fluctuation in the energies of the Kohn-Sham eigenvalues (at room temperature, RMS variations of *tenths* of an eV are observed, an effect far exceeding kT!). Localized eigenvectors conjugate to the fluctuating eigenvalues also show dramatic fluctuations[68]. From consideration of the Kubo formula, these "eigenvector fluctuations", which are strongly temperature dependent, will affect transport in the disordered environment. We have shown that the experimental thermal broadening of the electronic DOS observed in experiments on a-Si and an associated asymmetry in the temperature dependence of the Urbach decay energy for conduction and valence tails could be understood by studying the RMS fluctuations[49].

## Phonon-induced changes in electron states

To quantify the role of the electron-lattice coupling and motivate the rest of the proposed work, we begin by computing a deformation potential, which measures the response of a certain electronic energy eigenvalue to a particular phonon[48]. We choose a-Si for this purpose, though we have seen that the results are qualitatively similar to amorphous (carbon and selenium) and glassy (chalcogenide) systems. We consider an electronic eigenvalue, $\lambda_n$, say in one of the band tails in the amorphous material. To estimate the sensitivity of $\lambda_n$ to a coordinate distortion the Hellmann-Feynman theorem, gives $\partial\lambda_n/\partial\mathbf{R}_\alpha = \langle\psi_n|\partial H/\partial\mathbf{R}_\alpha|\psi_n\rangle$, and with the assumption of harmonic dynamics, leads easily to the electron[n]-phonon[$\omega$] coupling $\Xi_n(\omega)$:

$$\Xi_n(\omega) = \Sigma_\alpha \langle\psi_n|\partial H/\partial\mathbf{R}_\alpha|\psi_n\rangle \chi_\alpha(\omega).$$

$\Xi$ is easily computed as a byproduct of any *ab initio* calculation of the vibrational modes[48]. In this equation, R is the 3N vector of displacements, $\chi_\alpha(\omega)$ is a vibrational normal mode with displacement index $\alpha$ and frequency $\omega$.

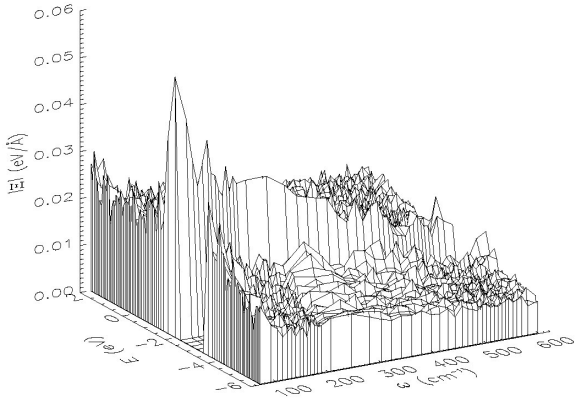

**Fig. 9. Electron-phonon coupling surface plot for 216 atom model of amorphous Si. Phonon frequency $\omega$, electron energy E and absolute value of electron-phonon coupling $\Xi$ (see text). The optical gap extends from –3.45 to –2.11 eV. Minimal basis FIREBALL calculation was used.**

In Fig. 9, we present the results of such a calculation for a WWW-type model of a-Si due to Djordjevic, Thorpe and Wooten[69], which has no coordination defects, though there are a small number of strained structures that lead to a highly realistic distribution of localized tail states. From Fig. 9 note that (1) the electron-phonon coupling is larger for conduction tail states than valence tail states (the conduction tails are also more localized), 2) the *low energy* phonons are more important to the tail states than optical phonons, and 3) the electron-phonon coupling falls off rapidly for electron energies away from band edges.

The large and intricately structured electron-phonon coupling surface illustrated in Fig. 9 arises from the special nature of localized states and is thus unique to systems with strong localization (amorphous materials and certain types of defects in crystals). The larger electron-phonon coupling for the conduction states has been indirectly seen in experiments on a-Si:H. Similar effects are seen in chalcogenide glasses..

## Solutions of the Time-Dependent Kohn-Sham (Schrödinger) Equation

It is clear that this picture is not an entirely "realistic" description of electron dynamics in the amorphous network. A more plausible picture is that at some initial time, an electron is in some (possibly localized) state. The motion of the lattice induces time-dependence in the electronic Hamiltonian, so that a state which is pure at $t=0$ will not remain so at later times. This has been named "phonon-induced delocalization" by Thomas[51], who has emphasized the significance of the successive delocalization with Anderson models. In other words, the electrons are scattered into successively less-localized (mixed) states by the lattice vibrations. On intuitive grounds, one would expect the intensity of phonon-induced transitions to depend strongly on (1) temperature, and (2) proximity of localized wave functions (both in energy and real space), since mixing is probable only when eigenvalues are close in energy to enable resonant tunneling.

To explicitly compute the time development of an electron packet we have integrated the time-dependent Schrödinger equation (TDSE):

$$i\hbar \partial \psi / \partial t = H\psi$$

Here $\psi$ is the wave function of a single electron and $H$ is the one electron (density functional) Hamiltonian matrix[70] for the host (here, a-Si). For sufficiently small step $\tau$ we use the Crank-Nicholson[71] scheme to generate an approximate time development operator:

$$U(\tau) = [1 + i\tau H/2\hbar]^{-1}[1 - i\tau H/2\hbar]$$

We make this choice of U, since this form is exactly unitary for any $\tau$. With the use of an appropriate (Löwdin) orthonormalization for the non-orthogonal basis[72], we solved the TDSE for 216-atom and 64-atom a-Si models. Two localized states, the highest occupied molecular orbital (HOMO) and the lowest unoccupied molecular orbital (LUMO), were chosen to study the delocalization process. These edge states are localized on several atoms in the 216-atom and 64-atom models and are either band tail or mid gap states as we discuss separately for each model. The electronic diffusion is monitored using the inverse participation ratio (IPR), which measures spatial locality of a state (large IPR means more compact state).

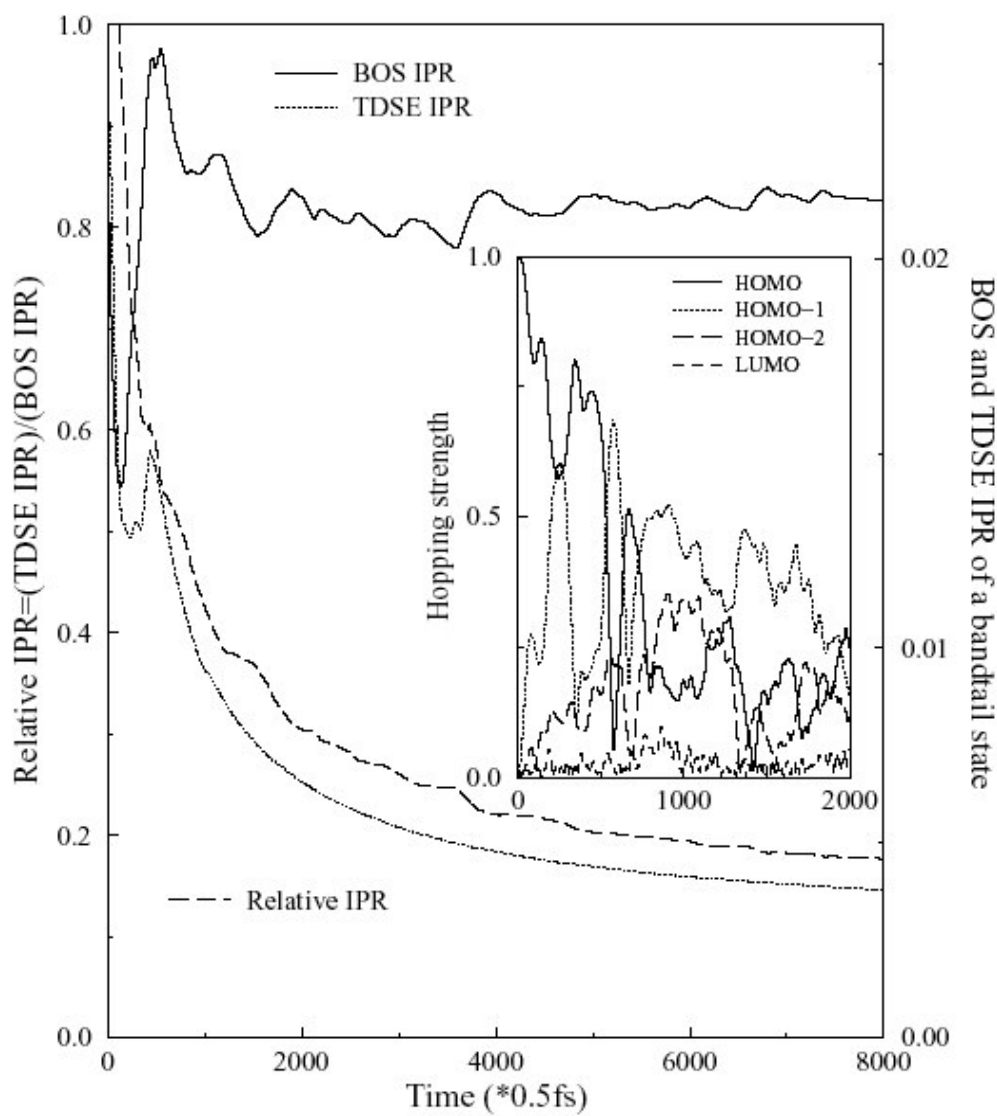

**FIGURE 10.** Time-development of localization in a-Si from time-dependent Schrödinger (Kohn-Sham) equation. Evolution of the localization (IPR) of a band tail state (HOMO) in 216 atom model of a-Si. "BOS" is time-averaged "Born-Oppenheimer snapshots (see text); lower two curves are from time dependent Kohn-Sham equation. Insert: spectral electron diffusion or leakage of the HOMO into adjacent energy states induced by thermal MD at 300K. "Relative IPR" is ratio of the TDSE calculation to the BOS result.

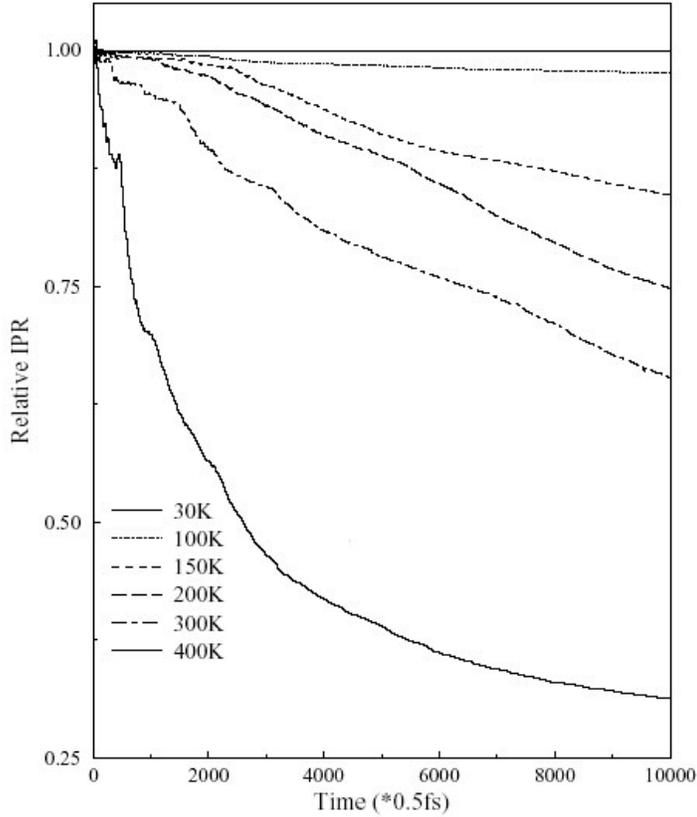

**FIGURE 11. Temperature dependence of relative localization of mid-gap state in the model.**

For present purposes, we summarize the following main results: (1) there is a qualitative difference between the "Born-Oppenheimer snapshots" (BOS) (time-averaged IPR from the instantaneous eigenstates obtained at each time step in the thermal MD simulation) and solutions of the time-dependent Kohn-Sham equation when there is diffusive behavior predicted by the TDSE equation; (2) there is strong temperature dependence in the hopping; (3) the hopping between localized states can be between *clusters* of atoms (the details of the structure of the clusters depends sensitively to energy); (4) Mott's variable range hopping is clearly seen, though the limitations of the model are apparent (especially the assumption of a spherical hopping volume), and (5) The hopping or electron diffusion is rapid if there are states nearby both in energy and in real space (for a spectrally isolated mid-gap state, the diffusion is much slower than for a band tail state, which occurs in a denser part of the density of states – compare Figures 10 and 11). The approach given here enables direct measure of thermally induced transfer from one localized state to another, as seen in the inset in Figure 10.

From a technical or methodological perspective, we believe that the preceding is a good beginning to addressing the questions about the time-dependence raised above since: (1) the disorder is realistic (arising from realistic structural models); (2) the thermal disorder (lattice motion) is also realistic (computed phonon DOS reproduce experiment); (3) the resulting behavior: temperature dependence, cluster hopping, and spectral electron diffusion seem reasonable.

## Acknowledgements

We thank the US National Science Foundation for support under grants DMR-0074624 and DMR-0205858 and DMR-0310933. We thank Professors Himanshu Jain and John Abelson for collaborations, and Dr. R. L. Cappelletti, and Dr. S. N. Taraskin for helpful discussions. Dr. J. C. Phillips gave helpful advice in connection with electron-phonon couplings and associated effects for localized states. Finally, we thank Professor N. Mousseau for providing us with the (much used) WWW cells of this paper.

of course smaller still.